\documentstyle[revtex]{aps}
\begin{document}
\begin{center}
\Large\bf

Supersensitive avalanche silicon drift photodetector.

\vspace{1.0cm}

\large

Z. Ya. Sadygov$^*$, M. K. Suleimanov,T. Yu. Bokova

\large\it
Joint Institute for Nuclear Research, Dubna,Russia

*) E-mail: sadygov@cv.jinr.ru

\vspace{1.0cm}

                           ABSTRACT
\end{center}

{  Physical principles of performance and main characteristics of a novel avalanche photodetector developed on the basis
   of MOS(metal-oxide-silicon) technology is presented. The photodetector contains a semitransparent gate electrode and a drain
   contact to provide a drift of multiplicated charge carriers along
   the semiconductor surface. A high gain(more than $10^4$) of
photocurrent was achived due to the local negative feedback effect realizied on
   the $Si-SiO_2$ boundary.

          Special attention is  paid  to  the  possibilities  of
   development of a supersensitive avalanche CCD (charge coupled
   device) for detection of individual photons in visible and ultraviolet
   spectral regions.  Experimental results obtained with  a two-element
   CCD prototype are discussed.}

\begin{center}

                        INTRODUCTION
\end{center}

 {       For last decades, researchers have sought solid-state
   alternatives to photomultiplier tubes(PMT) to be used for application
 in physical experiments and medical tomography. However, there is not
 yet any adequate solid-state analog comparable to commercial PMT to
 detect week light pulses consisting of a few photons at room
 temperature.  During last years, we have investigated possibilities to
    made relatively cheap APDs on the basis of MRS(metal-resistive
 layer- semiconductor) structures fabricated on low resistive silicon
 wafers [1,2].  The main peculiarity of the MRS APDs is a local negative
 feedback (LNF) effect which results in a self-stabilized avalanche
 process in the semiconductor. These LNF APDs have been developed for
 use in red and infrared spectrum regions.  Recently, we reported some
        characteristics of another design of LNF APD's having a high
 sensitivity in visible and ultraviolet spectral regions [3]. This
 report is devoted to physical principles of performance and main
 characteristics of a novel design of the LNF APD.}

\begin{center}
            DEVICE DESIGN AND PRINCIPLES OF OPERATION
\end{center}

{ The new LNF APD contains a semitransparent titanium layer
 (gate electrode) separated from the semiconductor surface
 by a silicon oxide layer and a guard-ring(drain electrode) to provide
  a drift of multiplicated charge carriers along the $Si-SiO_2$ boundary
 (see Fig.1). The titanium gate and drain electrode can be provided
 with individual or common(key $K_1$ is closed) aluminum contacts for
 voltage supply.

          Avalanche multiplication of photocurrent occurs in the p-n
 junction(under the gate electrode) where the breakdown potential is
 specially reduced due to an additional ion-implantation. The voltage applied to
 the LNF APD is distributed between the depletion layer of the
 semiconductor and the oxide layer. The hole curriers to
 be caused by avalanche process in a given microregion of the
 p-n junction is collected in a small area of the $Si-SiO_2$ boundary,
 reducing the value of voltage drop in this microregion of the semiconductor.
 As a result each start avalanche process is self-switched in
 a few nanoseconds in the given microregion. A drift of the hole charge curriers from
 the avalanche region to the drain contact is provided by a high resistive
 layer during about 100 ns after switching of the given analanche process.
The parameters(thickness and surface resistivity) of this resistive layer are field-dependent, and so the LNF effect
 can be adjusted by the gate potential. Such character of avalanche  process is called the local negative feedback effect.}

\begin{center}

             MAIN PECULIARITIES OF DEVICE PERFORMANCE
\end{center}

    {    Fig.2 shows the measured gain of photocurrent versus negative
 bias on the drain electrode. The LNF APD sample had a photosensitive area
 of 1mm diameter. An initial pulsed photocurrent with a duration
 of 0.4ms and an amplitude of 6nA was used in these measurments.
 One can see that the character of avalanche process in the LNF APD
 is fully defined by the gate potential. At fixed gate biases,
 a situation can be achieved where gain becomes independent of drain
 potential. This peculiarity of the tested device indicates the unique
 possibility of building multichannel avalanche photodetectors with a high
 spatial uniformity of gain.

       Another type of operation of this device is connected with a pulsed
 third level supply of the gate electrode(for example, $V_g$=-69.5V,
 -55V, -45V at $V_d$=-63V=const). Under these conditions, a CCD(charge
 coupled device) type performance of the LNF APD was realized. A high
 gain of about $10^4$ was obtained. A light emitted diode with a
 wavelength of 450nm and a pulse duration of 50ns was used as a signal
 for detection.

       The amplitude distributions of the signals detected by the LNF APD
 are presented in Fig.3 for light pulses which contain one photon in
 average. The photomultiplier FEU-130(made in Russia) was used to count
 the average number of photons in light pulses. One can see that a single
 photon detection mode with a high efficiency was realized at room
 temperature. If we take the threshold amplitude equal to the 150-th
 channel, then the detection efficiency of the LNF APD is about 25$\%$.}

\begin{center}
                     PERSPECTIVES
\end{center}

    {  As shown above there are real possibilities of development
 of a super sensitive multichannel photodetectors as well as
 an avalanche type CCD for single photon detection in visible
 and ultraviolet spectrum regions. To our mind, these unique properties
 of the presented LNF APDs  would be good prospects for future
 applications in high energy physics, high speed photonics and
 medical tomography.}

\begin{center}

                       REFERENCES

\end{center}

1.Z.Sadygov et.al. Proc. SPIE- The Intern. Soc. for Opt. Eng., 1621,
  (1991), 158-168.

2.Z.Sadygov et.al. IEEE Trans. Nucl.Sci., 43, 3, (1996), 1009-1013.

3.Z.Sadygov. Russian patent N2086027, application of May 30-th, 1996.

4.N.Bacchetta, Z.Sadygov et.al.. Nucl.Instr.and Meth, A387, (1997),
  225-230.

\begin{center}

FIGURE CAPTIONS
\end{center}

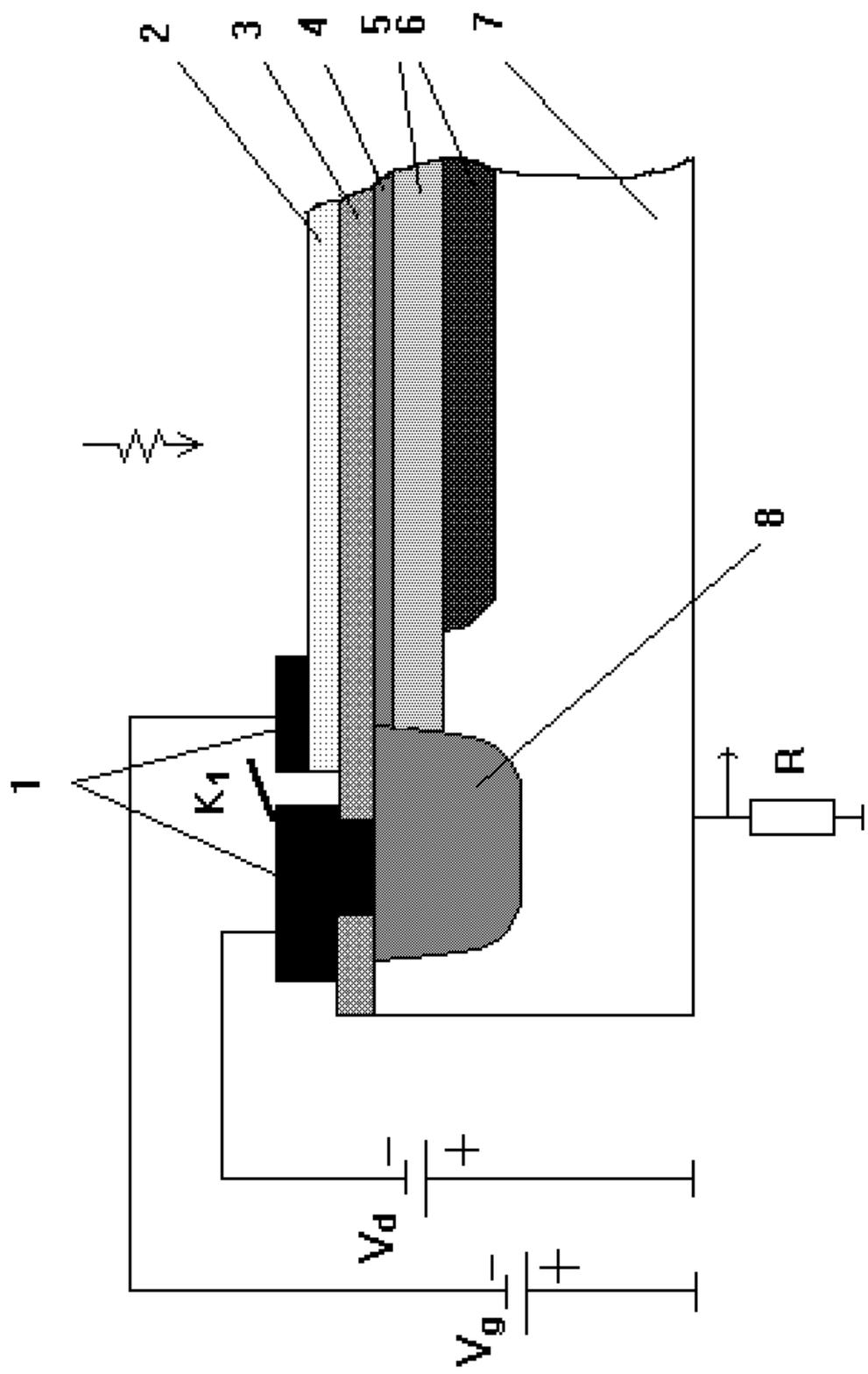
    Figure 1. Simplified structure of the LNF APD. 1- thick Al electrodes;
              2- semitransparent Ti gate electrode; 3- dielectric layer
              ($SiO_2$); 4-surface drift layer;  5- p-Si layer; 6- n-Si
layer with additional ion-implantation; 7- n-Si wafer; 8- p-Si guard (drain)
              ring.

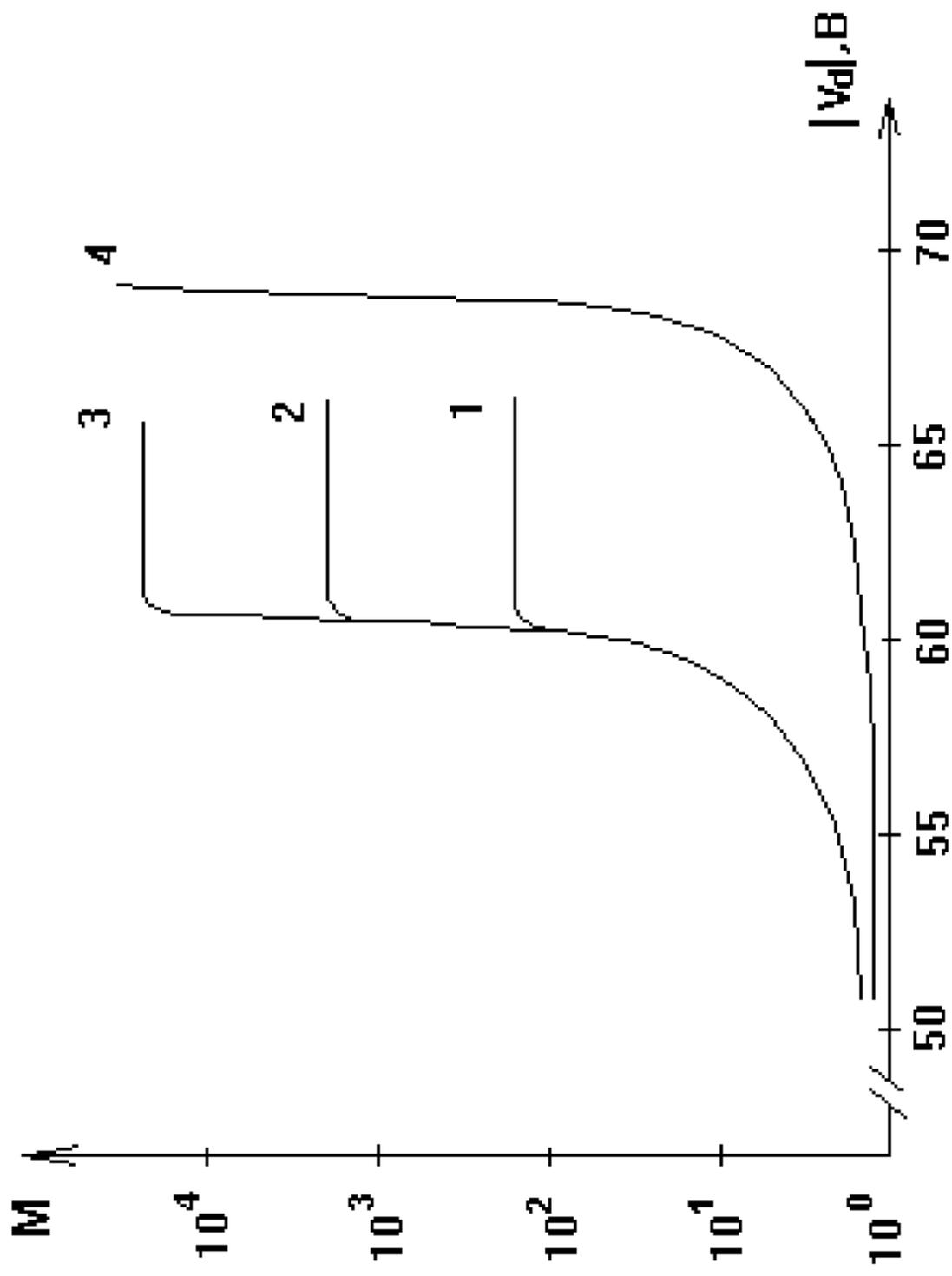
   Figure 2. Photocurrent gain as a function of drain voltage.
             1- $V_g$=-68.5V=const; 2- $V_g$=-69.0V=const; 3-
              $V_g$=-69.5V= const; 4- $V_g$=$V_d$ (key $K_1$ is closed).

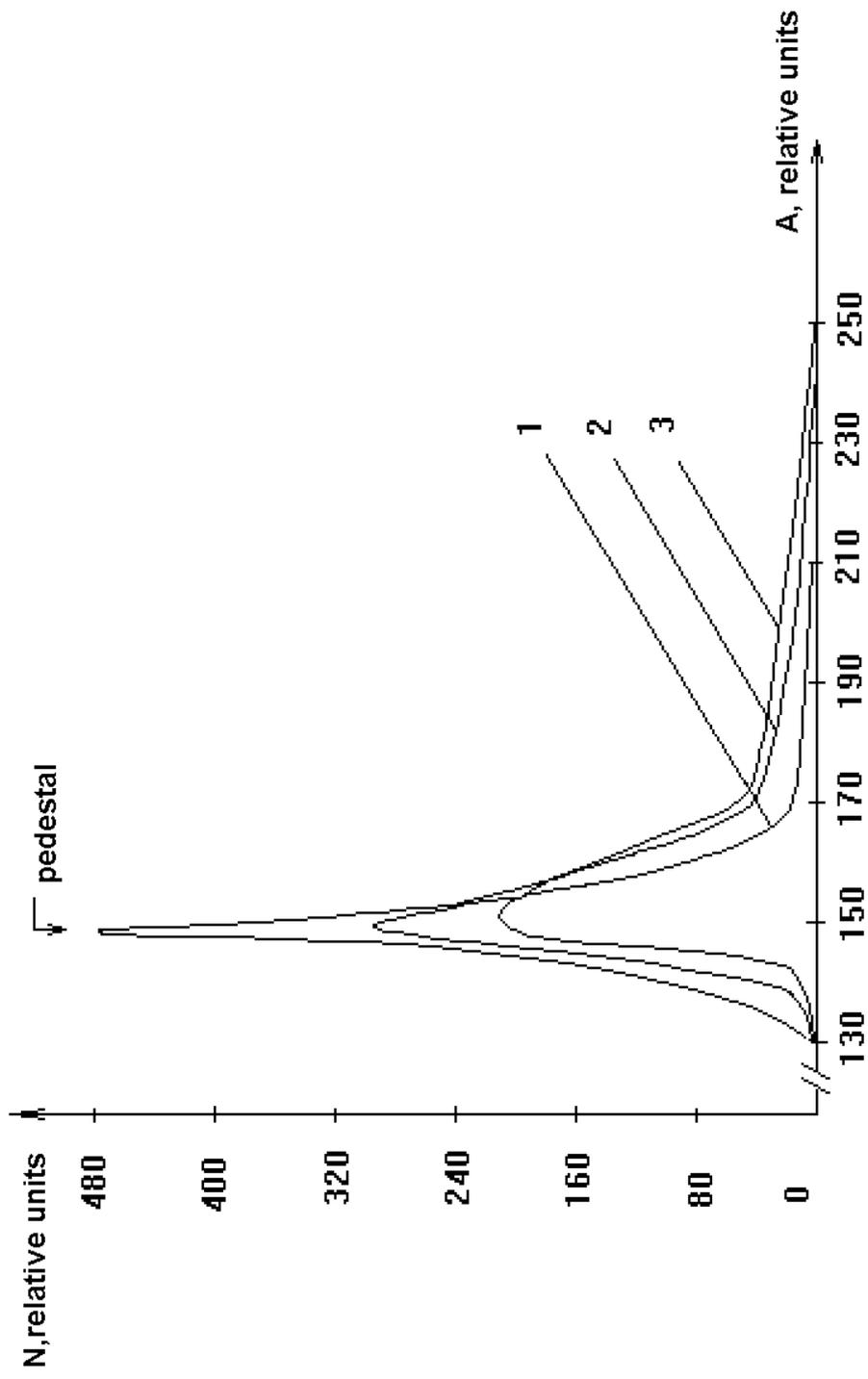
   Figure 3. Amplitude spectrum of LNF APD output pulses.
            1- dark condition; 2- single-photon mode;
            3- double-photon mode

\end{document}